\documentclass[12pt,epsf]{aastex}
\usepackage{psfig,rotate,amsmath}
\title{Recent Torque Reversal of 4U 1907+09}
\author{S.\c{C}. \.{I}nam$^1$, \c{S}. \c{S}ahiner$^2$, A. Baykal$^2$  \\
$^1$ Department of Electrical and
Electronics Engineering,\\ 
 Ba\c{s}kent University, 06530 Ankara, Turkey \\ inam$@$baskent.edu.tr \\
$^2$ Physics Department,\\ Middle East Technical University, 06531 Ankara,
Turkey \\ seyda$@$astroa.physics.metu.edu.tr, altan$@$astroa.physics.metu.edu.tr }

\date{}

\begin{document}

\begin{abstract}

We present timing and spectral analysis of RXTE-PCA observations of the accretion powered pulsar 4U 1907+09 between June 2007 and August 2008. 4U 1907+09 had been in a spin-down episode with a spin-down rate of $-3.54\times10^{-14}$ Hz s$^{-1}$ before 1999. From RXTE observations after March 2001, the source showed a $\sim 60$\% decrease in spin-down magnitude and INTEGRAL observations after March 2003 showed that source started to spin-up. We found that the source recently entered a new spin-down episode with a spin-down rate of $-3.59 \times 10^{-14}$ Hz s$^{-1}$. This spin-down rate is pretty close to the previous long term spin-down rate of the source measured before 1999. From the spectral analysis, we showed that Hydrogen column density varies with the orbital phase. 

{\bf{Keywords:}} X-rays: binaries, pulsars:individual:4U 1907+09, stars:neutron, accretion, accretion discs 
\end{abstract}

\maketitle
\section{Introduction}

4U 1907+09 is a High Mass X-ray Binary (HMXB) system discovered by the Uhuru survey (Giacconi et al. 1971). The system consists of an X-ray pulsar accreting mass from a blue supergiant companion star. An orbital period of 8.38 days was found from Ariel V observations (Marshall \& Ricketts 1980). Makishima et al. (1984) found the spin period value of 437.5 s using Tenma observations. Eccentricity of the orbit of the system was found to be $\sim$0.16 from EXOSAT observations (Cook \& Page 1987). From RXTE observations, In't Zand et al. (1998) found revised orbital parameters of the system. The evidence of two phase-locked flares at the folded orbital profile was reported (Marshall \& Ricketts, 1980; In't Zand et al. 1998). The presence of the two flares per orbit leaded to the suggestion of an equatorial disk like envelope around a Be type stellar companion (Cook \& Page 1987). Although  the variation in H$\alpha$ emission in the optical observations strengthened the suggestion of a Be star (Iye 1986), Cox et al. (2005) suggested the possibility of an OB supergiant companion previously mentioned by Schwartz et al. (1980) and Van Kerkwijk et al. (1989). Cox et al. (2005) identified  the companion as an O8-09 Ia supergiant with a distance of $\sim$5 kpc. 

4U 1907+09 had been steadily spinning down for more than 15 years with an average rate of $\dot \nu=-3.54\times 10^{-14}$ Hz s$^{-1}$ (Cook \& Page 1987; in't Zand et al. 1998; Baykal et al. 2001; Mukerjee et al. 2001). Afterwards, RXTE observations in 2001 showed $\sim$60\% decrease in the magnitude of the spin down rate (Baykal et al. 2006). It has recently been reported that the pulse period evolution has shown a torque reversal with $2.58\times 10^{-14}$ Hz s$^{-1}$, confirming a spin up episode after March 2003 (Fritz et al. 2006). 
 
Spectral studies (Schwartz et al. 1980; Marshall \& Ricketts 1980; Makishima et al. 1984; Cook \& Page 1987; Chitnis et al. 1993; in't Zand et al. 1997; Coburn et al. 2002) on 4U 1907+09 indicate a highly variable hydrogen column density over the binary orbit, between $\sim1\times10^{22}$ cm$^{-2}$ and $\sim9\times10^{22}$ cm$^{-2}$. The power law photon index is between $\sim$0.8 and $\sim$1.5 with a high energy cutoff around $\sim$13keV. Extensive energy spectra generated from Ginga (Mihara 1995; Makishima et al. 1999) and BeppoSAX (Cusumano et al. 1998) observations exhibit absorption features at $\sim$19keV and $\sim$39keV referring to the fundamental and the second harmonic of a cyclotron absorption feature corresponding to a surface magnetic field strength $2.1\times10^{12}$ G. 

In this paper, we present timing and spectral analysis of RXTE monitoring observations of 4U 1907+09 between  MJD 54280 and MJD 54690. First, we provide a brief description of observation and analysis procedure in Section 2. The pulse timing analysis, indicating a new spin-down episode, is reported in Section 3. In Section 4, we present the results of X-ray spectral analysis.  

\section{Observations and Data Analysis}

The analysis of RXTE monitoring observations of 4U 1907+09 is presented in this paper. The observations between June 2007 and August 2008 are analysed, each having an average exposure time $\sim$2 ksec (see Figure 1). The data products of Proportional Counter Array (PCA) onboard RXTE have been reduced for the analysis. 
RXTE-PCA is a pointed instrument (Jahoda et al. 1996); it consists of five co-aligned identical proportional counter units (PCUs) sensitive in the energy range 2-60 keV. The effective area of each detector is approximately $\sim$1300cm$^2$ and the energy resolution is $18\%$ at 6 keV. A field of view (FOV) of 1$^{\circ}$ at full width at half maximum (FWHM) is small enough that, with the large geometric area, PCA provides a highly sensitive time resolution. Operations are carried on with various PCUs turned off during observations in order to extend the life time of the instrument. The number of running PCUs during the observations of 4U 1907+09 varies between one and three. PCU0 and PCU1 have increased background levels due to the loss of their propane layers. As it is not recommended to use data of these PCUs for spectral analysis, data of PCU2 were used during spectral analysis. No PCU selection is done for the timing analysis, since propane losses do not affect high resolution timing. 

The standard RXTE analysis software HEASOFT 6.4 is used for data reduction. The Standard2f mode data with 128 energy channels are examined for the spectral analysis. Since the timing resolution of Standard2f is low, the GoodXenon mode data are used for the timing analysis using 1s long binning. The filtering applied to the data is comprised of excluding the times when, elevation angle is less than 10$^{\circ}$ and offset from the source is greater than 0.02. Electron contamination of PCU2 is also confirmed to be less than 0.1. The latest PCA background estimator models supplied by the RXTE Guest Observer Facility (GOF), Epoch 5C are used to generate the background light curves and spectra. The model estimation is based on the rate of very large solar events, spacecraft activation and cosmic X-ray emission. Since 4U 1907+09 is near the Galactic plane and the supernova remnant W49B, additional background estimation is done for X-ray spectral analysis (see Spectral Analysis section).

\begin{figure}[tb]
\begin{center}
\hspace{0.5cm}
\psfig{file=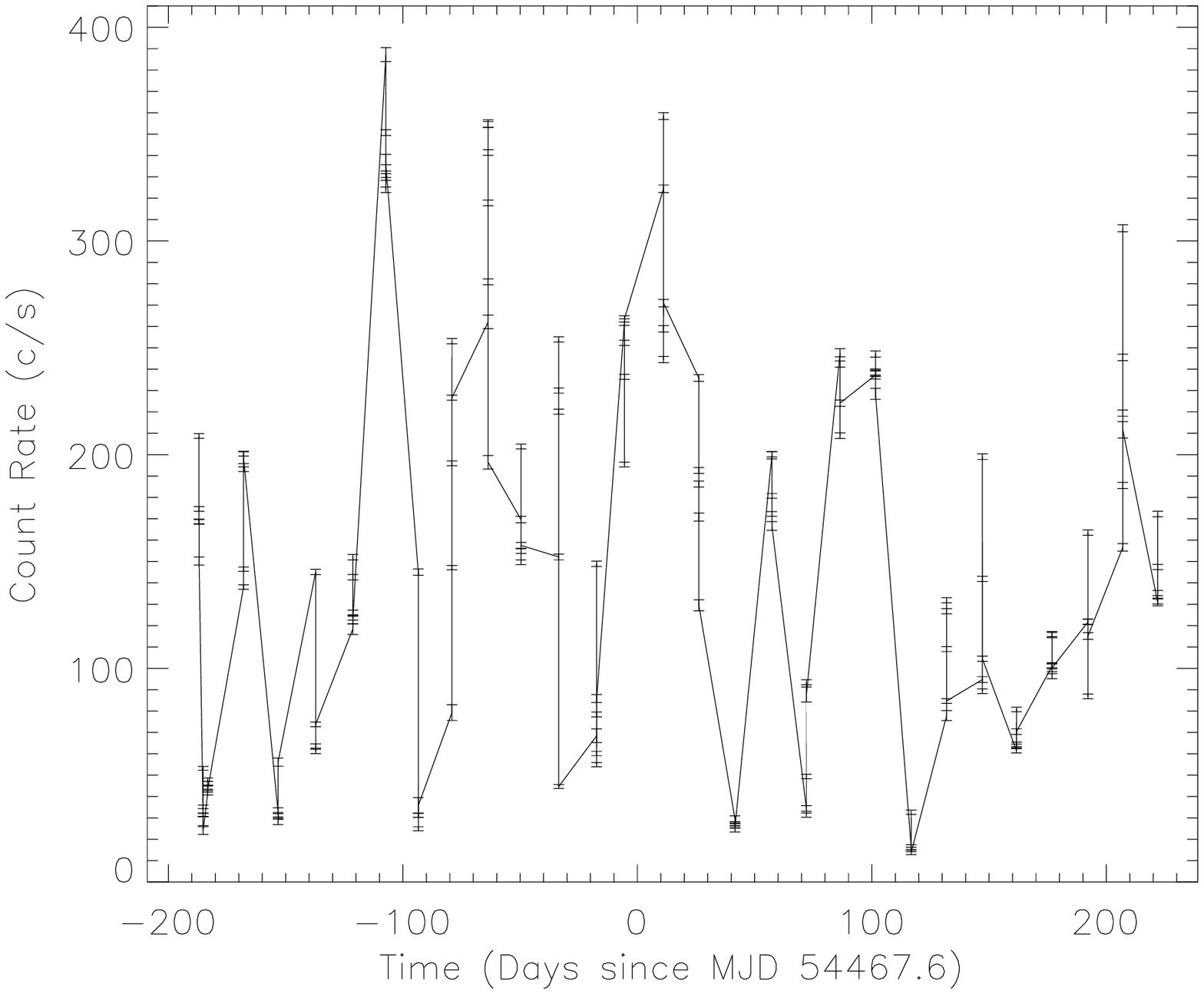,height=10cm,width=11.5cm}
\end{center}
\begin{center}
\small{Fig. 1 -- 440s binned lightcurve of 4U 1907+09}
\end{center}
\end{figure}

\section{Timing Analysis}

For timing analysis, the background subtracted light curves were corrected to the barycenter of the solar system and to the binary orbital motion of 4U 1907+09. For the correction of binary orbital motion, we used the orbital parameters deduced by in't Zand et al. 1998. We folded the $\sim$2 ksec long light curve segments outside intensity dips on statistically independent trial periods (Leahy et al. 1983). A template pulse profile was obtained from the first three observations with a total exposure of $\sim$6 ksec by folding the data on the period giving the maximum $\chi^2$. These three observations were made within the first week after the observations started.  Template pulse profile was analytically represented by its Fourier harmonics (Deeter \& Boynton 1985) and cross-correlated with the harmonic representation of average pulse profiles from each $\sim$2 ksec long observation. The pulse phases calculated from this analysis are presented in Figure 2.  

\begin{figure}[tb]
\begin{center}
\hspace{0.5cm}
\psfig{file=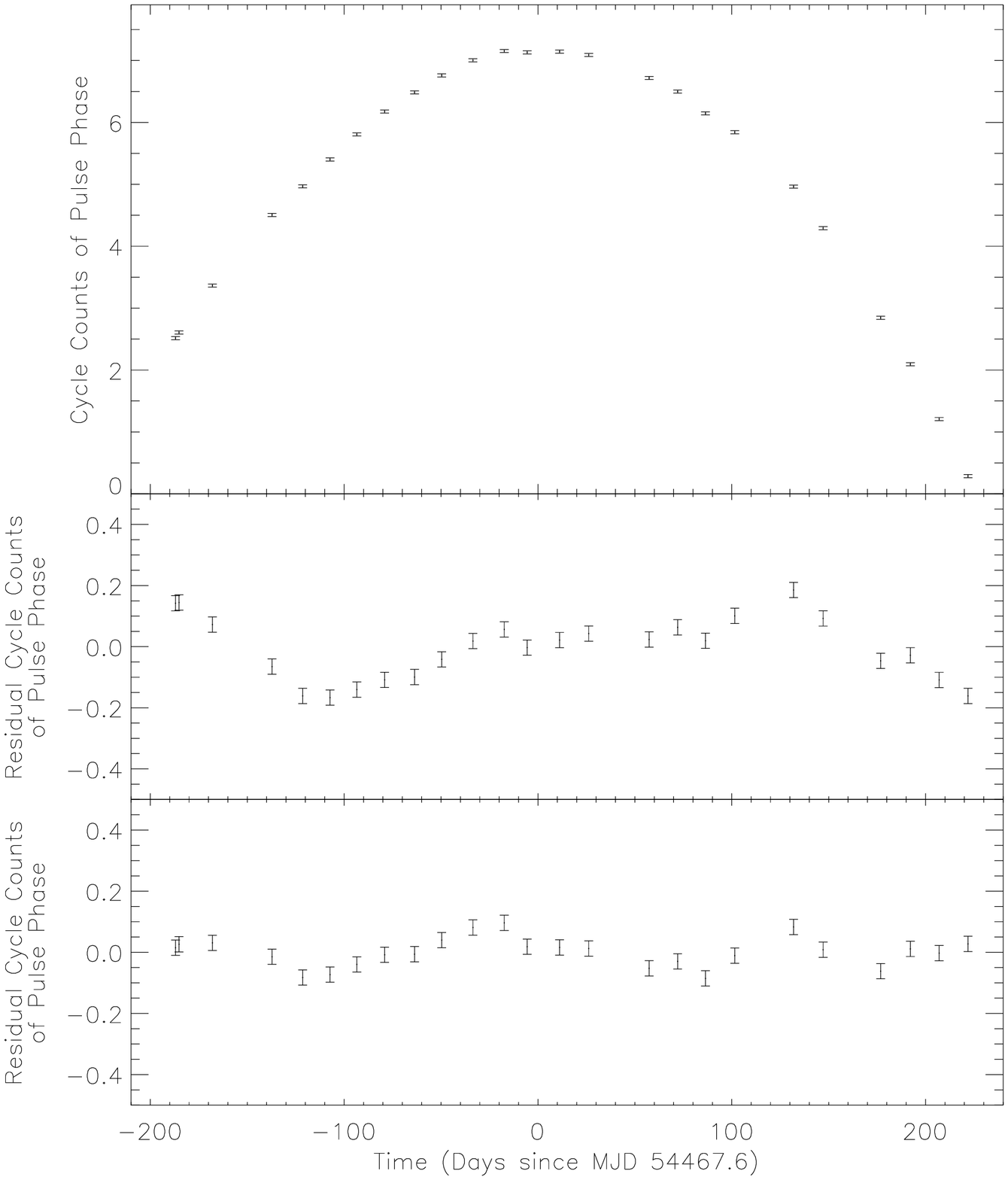,height=11cm,width=11cm}
\end{center}
\begin{center}
\small{Fig. 2 -- Pulse phase and its residuals after the removal of quadratic (middle panel) and cubic (lower panel) polynomials.}
\end{center}
\end{figure}

To estimate pulse frequency derivatives, the pulse phases were fitted to the polynomial
\begin{equation}
\delta \phi = \phi_{o} + \delta \nu (t-t_{o})
+ \frac{1}{2} \dot \nu (t-t_{o})^{2} +\frac{1}{6} \ddot \nu (t-t_{o})^{3}
\end{equation}
where $\delta \phi $ is the pulse phase offset deduced from the pulse
timing analysis, $t_{o}$ is the mid-time of the observation, $\phi_{o}$ is
the phase offset at t$_{o}$, $\delta \nu$ is the deviation from the mean
pulse frequency (or additive correction to the pulse frequency), $\dot
\nu $ and $\ddot \nu$ are the first and second pulse frequency derivatives of the source.
The pulse phases and
the residuals of the fit after the removal of the quadratic (without the cubic term in Equation 1) and cubic polynomial (Equation 1)
are also presented in Figure 2. Timing solution of the source is presented in Table 1.

\begin{table}[bt]
\caption{Timing Solution of 4U 1907+09}

\begin{center}
\begin{tabular}{l | c} \hline
Parameter &  Value  \\ \hline
Epoch & MJD 54467.6(1) Days \\ 
Orbital Period & 8.3753(1)   \\ 
Projected Semi-major Axis (lt s) & 83(2) \\
Eccentricity & 0.28(4) \\
Longitude of Periastron & 330(7) \\
Spin Frequency($\nu$) & $2.266460(2)\times 10^{-3}$ Hz \\
Spin Frequency Derivative ($\dot \nu$) & $-3.59(2) \times 10^{-14}$ Hz s$^{-1}$ \\
$\ddot \nu$ & $-5.3(4)\times 10^{-22}$ Hz s$^{-2}$ 
\\ \hline
\end{tabular}
\end{center}
\end{table}

The pulse frequency derivative from the sequence of 25 pulse phases from the observations between June 2007 and August 2008
was measured as $-3.59(2) \times 10^{-14}$ Hz s$^{-1}$. This measurement shows that the source started to spin-down with a spin-down rate close to long term spin-down rate of the source (Baykal et al. 2001).

\begin{figure}[tbf]
\begin{center}
\hspace{0.5cm}
\psfig{file=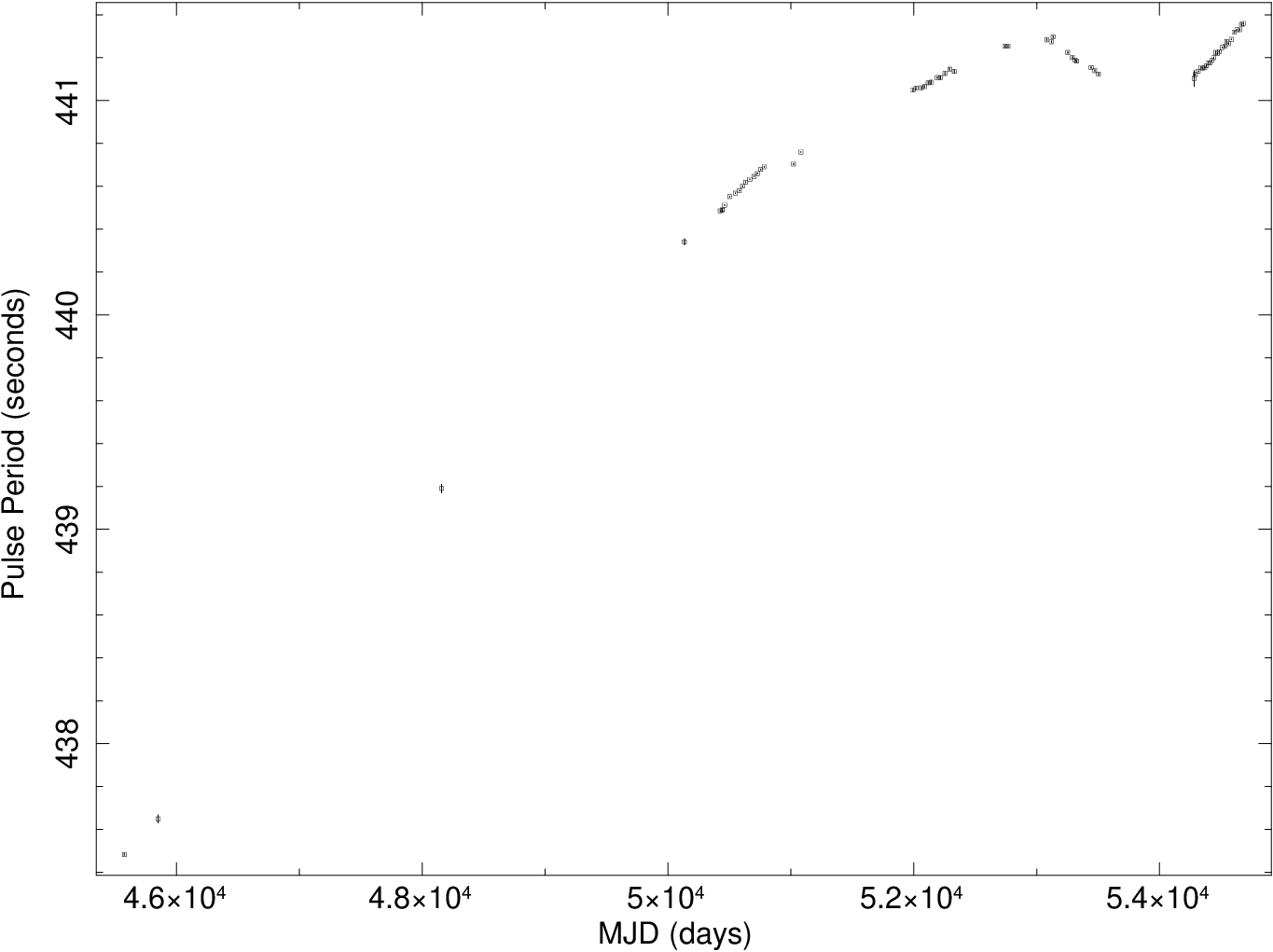,height=12cm,width=13.5cm,angle=-90}
\end{center}
\begin{center}
\small{Fig. 3 -- Pulse period measurements of 4U 1907+09. Recent spin-down trend of the source is evident between MJD 54281 and MJD 54682.}
\end{center}
\end{figure}

To demonstrate the torque reversal and show the recent spin-down trend of the source, we estimated the pulse frequency histories by taking the derivatives of each pairs of pulse phases from the pulse phases of this work. We presented pulse frequency history of the source in Figure 3, and listed all pulse period measurements in Tables 2.

\begin{table}[p]
\caption{Pulse period measurements of 4U 1907+09}
\scriptsize{\begin{center}
\begin{tabular}{|c c c || c c c|}\hline
Epoch & Pulse Period & Reference & Epoch & Pulse Period & Reference\\
(MJD) & (s) &   & (MJD) & (s) &    \\ \hline \hline
45576 & 437.483$\pm$0.004 & Makishima et al.1984   & 53121.1 & 441.274$\pm$0.005 & Fritz et al. 2006 \\
45850 & 437.649$\pm$0.019 & Cook \& Page 1987      & 53133.4 & 441.297$\pm$0.005 & Fritz et al. 2006 \\
48156.6 & 439.19$\pm$0.02 & Mihara 1995  	   & 53253.6 & 441.224$\pm$0.010 & Fritz et al. 2006 \\  
50134 & 440.341$\pm$0.014 & In't Zand et al.1998   & 53291.3 & 441.201$\pm$0.005 & Fritz et al. 2006 \\
50424.3 & 440.4854$\pm$0.0109 & Baykal et al.2006  & 53314.0 & 441.188$\pm$0.005 & Fritz et al. 2006 \\
50440.4 & 440.4877$\pm$0.0085 & Baykal et al.2001  & 53324.7 & 441.183$\pm$0.005 & Fritz et al. 2006 \\
50460.9 & 440.5116$\pm$0.0075 & Baykal et al.2006  & 53443.4 & 441.154$\pm$0.005 & Fritz et al. 2006 \\
50502.1 & 440.5518$\pm$0.0053 & Baykal et al.2006  & 53473.3 & 441.139$\pm$0.005 & Fritz et al. 2006 \\
50547.1 & 440.5681$\pm$0.0064 & Baykal et al.2006  & 53503.8 & 441.124$\pm$0.005 & Fritz et al. 2006 \\
50581.1 & 440.5794$\pm$0.0097 & Baykal et al.2006  & 54281.5  &      441.1030$\pm$0.0372 & This work \\
50606.0 & 440.6003$\pm$0.0115 & Baykal et al.2006  & 54291.0  &      441.1213$\pm$0.0038 & This work \\					      
50631.9 & 440.6189$\pm$0.0089 & Baykal et al.2006  & 54315.0  &      441.1367$\pm$0.0021 & This work \\					      
50665.5 & 440.6323$\pm$0.0069 & Baykal et al.2006  & 54338.2  &      441.1545$\pm$0.0041 & This work \\					      
50699.4 & 440.6460$\pm$0.0087 & Baykal et al.2006  & 54353.3  &      441.1509$\pm$0.0046 & This work \\		
50726.8 & 440.6595$\pm$0.0105 & Baykal et al.2006  & 54367.3  &      441.1543$\pm$0.0047 & This work \\
50754.1 & 440.6785$\pm$0.0088 & Baykal et al.2006  & 54381.3  &      441.1623$\pm$0.0046 & This work \\
50782.5 & 440.6910$\pm$0.0097 & Baykal et al.2006  & 54396.2  &      441.1750$\pm$0.0042 & This work \\
51021.9 & 440.7045$\pm$0.0032 & Baykal et al.2001  & 54410.9  &      441.1761$\pm$0.0047 & This work \\
51080.9 & 440.7598$\pm$0.0010 & Baykal et al.2001  & 54426.0  &      441.1862$\pm$0.0041 & This work \\
51993.8 & 441.0484$\pm$0.0072 & Baykal et al.2006  & 54442.1  &      441.1992$\pm$0.0040 & This work \\
52016.8 & 441.0583$\pm$0.0071 & Baykal et al.2006  & 54456.1  &      441.2245$\pm$0.0056 & This work \\
52061.5 & 441.0595$\pm$0.0063 & Baykal et al.2006  & 54470.4  &      441.2185$\pm$0.0039 & This work \\
52088.0 & 441.0650$\pm$0.0063 & Baykal et al.2006  & 54486.3  &      441.2284$\pm$0.0043 & This work \\
52117.4 & 441.0821$\pm$0.0062 & Baykal et al.2006  & 54509.4  &      441.2472$\pm$0.0021 & This work \\
52141.2 & 441.0853$\pm$0.0082 & Baykal et al.2006  & 54532.3  &      441.2537$\pm$0.0044 & This work \\
52191.4 & 441.1067$\pm$0.0046 & Baykal et al.2006  & 54546.8  &      441.2756$\pm$0.0046 & This work \\
52217.2 & 441.1072$\pm$0.0077 & Baykal et al.2006  & 54561.6  &      441.2657$\pm$0.0043 & This work \\
52254.3 & 441.1259$\pm$0.0074 & Baykal et al.2006  & 54584.3  &      441.2855$\pm$0.0022 & This work \\
52292.0 & 441.1468$\pm$0.0065 & Baykal et al.2006  & 54607.1  &      441.3195$\pm$0.0043 & This work \\
52328.8 & 441.1353$\pm$0.0090 & Baykal et al.2006  & 54629.6  &      441.3301$\pm$0.0022 & This work \\
52739.3 & 441.253$\pm$0.005 & Fritz et al. 2006    & 54652.1  &      441.3307$\pm$0.0043 & This work \\
52767.1 & 441.253$\pm$0.005 & Fritz et al. 2006    & 54667.2  &      441.3549$\pm$0.0044 & This work \\
53083.9 & 441.283$\pm$0.005 & Fritz et al. 2006    & 54682.1  &      441.3596$\pm$0.0044 & This work \\

\hline \hline
\end{tabular}
\end{center}}
\end{table}

\section{Spectral Analysis}

\begin{figure}[tbh]
\begin{center}
\hspace{0.5cm}
\psfig{file=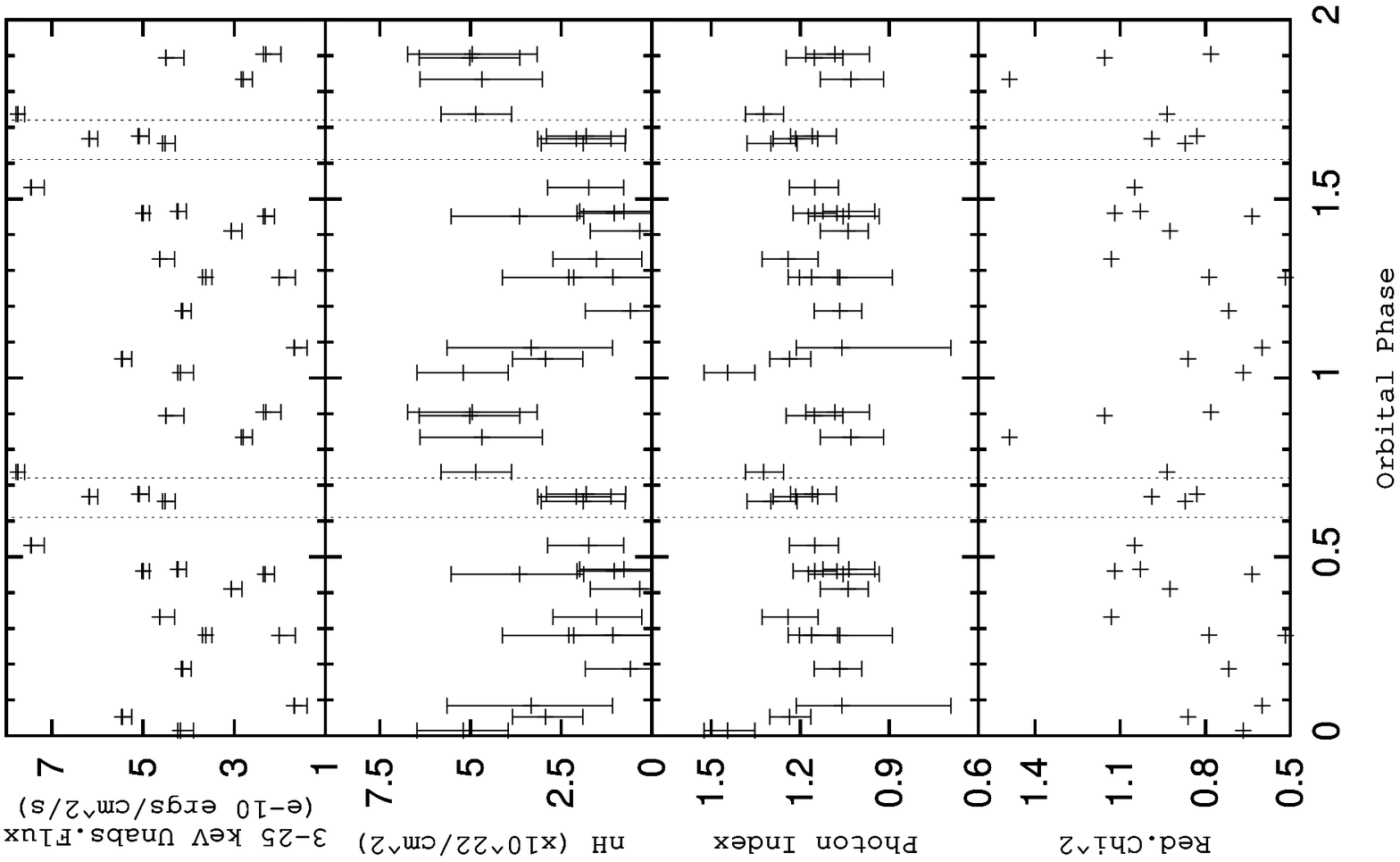,height=12cm,width=10cm,angle=-90}
\end{center}
\begin{center}
\small{Fig. 5 -- Orbital variations of the spectral parameters are plotted. Errors are calculated for 68\% confidence level. The dashed vertical lines indicate the time of periastron passage within 1$\sigma$.}
\end{center}
\end{figure}

The energy range of the spectral analysis was restricted to 3-25 keV range since PCA count statistics is poor beyond 25 keV. Along with the basic estimation of PCA background models, additional background model was needed because 4U 1907+09 is near the Galactic plane and the supernova remnant W49B. During the dip states of 4U 1907+09, no pulsed emission was observed and the count rate was consistent with the diffuse emission from the Galactic ridge (in't Zand et al. 1997). About 20\% of the processed RXTE observations were detected to be in dip state, so an overall dip state spectrum was constructed for comparison. It was found that the spectral parameters are consistent with the models of diffuse emission (Valinia \& Marshall 1998). Model estimated background was directly subtracted from each individual observation and the overall dip state spectrum was used as background in XSPEC. Systematic error of $2\%$ is applied during analysis to account for the uncertainties in the response matrix produced with PCARSP.
3-25 keV spectra of each individual non-dip observation were first modeled with power law with a high energy cutoff and photoelectric absorption. Deviation of residuals around 20 keV confirmed the presence of cyclotron absorption feature reported by Mihara (1995), Makishima et al. (1999) and Cusumano et al. (1998).  Among the several models tested to fit the fundamental cyclotron line at $\sim$19 keV, best results were obtained by cyclotron absorption line model "cyclabs" (Makishima et al. 1990). We first tried to fix the line energy to the reported value 18.9 keV, but better statistics is achieved when it is set free. The line energies was found to vary between $\sim$15.0 keV and $\sim$19.4 keV including 1$\sigma$ errors, which are consistent with the previous studies. 

The supernova remnant W49B has a strong emission feature at $\sim$6.4 keV (Miceli et al. 2006), but after a proper background subtraction, we did not find a significant iron line complex feature from 4U 1907+09. Although average spectrum (see Figure 4) has deviating residuals around 7 keV which may correspond to iron line, adding a Gaussian compenent to the spectral model did not improve the fit.

In order to investigate  orbital phase dependence of orbital variations of spectral parameters and X-ray flux, we present orbital phase resolved spectroscopy of the source in Figure 565. Sample spectral parameters corresponding to different orbital phases are listed in Table 3. 

\begin{table}[th]
\caption{Sample Spectral Parameters of 4U 1907+09}
\scriptsize{
\begin{center}
\begin{tabular}{|l | c c c c |} \hline
Observation ID &  93036-01-29-00 & 93036-01-04-00 & 93036-01-16-00 & 93036-01-10-00  \\ 
Observation Mid-Time (MJD) & 54675 & 54300 & 54479 & 54389 \\
Orbital Phase & 0.05 & 0.28 & 0.67 & 0.89 \\
\hline
$n_H$ ($10^{22}$ cm$^{-2}$) & $2.94_{-1.03}^{+0.91}$ & $1.08_{-1.08}^{+1.21}$ & $2.09_{-0.96}^{+1.06}$ & $5.02_{-1.38}^{+1.39}$ \\ 
Power Law Photon Index & $1.24_{-0.07}^{+0.07}$ & $1.16_{-0.09}^{+0.08}$ & $1.22_{-0.07}^{+0.08}$ & $1.15_{-0.10}^{+0.10}$ \\
Power Law Normalization (10$^{-2}$cts cm$^{-2}$ s$^{-1}$) & $3.90_{-0.59}^{+0.65}$ & $2.19_{-0.39}^{+0.45}$ & $4.00_{-0.60}^{+0.75}$ & $2.64_{-0.53}^{+0.67}$ \\
Cut-off Energy (keV) &  $12.26_{-0.35}^{+0.82}$ & $12.74_{-0.44}^{+1.01}$ & $12.88_{-0.74}^{+0.90}$ & $12.61_{-0.74}^{+0.98}$ \\
E-folding Energy (keV) & $7.67_{-0.78}^{+1.54}$ & $7.27_{-1.26}^{+1.96}$ & $10.88_{-1.73}^{+2.49}$ & $8.78_{-2.19}^{+3.89}$ 
\\
Cyclotron Depth & $0.41_{-0.19}^{+0.13}$ & $0.47_{-0.24}^{+0.23}$ & $0.43_{-0.13}^{+0.13}$ & $0.37_{-0.13}^{+0.30}$ \\
Cyclotron Energy (keV) & $16.74_{-0.59}^{+0.65}$ & $16.70_{-0.76}^{+0.78}$ & $16.63_{-0.58}^{+0.58}$ & $17.17_{-1.66}^{+1.23}$ \\
Cyclotron Width (keV) & $1.00_{-1.00}^{+1.91}$ & $1.14_{-1.14}^{+1.47}$ & $1.95_{-0.95}^{+0.98}$ & $2.53_{-2.53}^{+1.79}$ \\ 3-25 keV Unabsorbed Flux (10$^{-10}$ erg cm$^{-2}$ s$^{-1}$) & $5.46_{-0.21}^{+0.02}$ & $3.62_{-0.14}^{+0.08}$ & $6.18_{-0.18}^{+0.01}$ & $4.50_{-0.22}^{+0.02}$ \\
Reduced $\chi^2$ (41 d.o.f) & 0.86 & 0.79 & 0.99 & 1.16 \\

\hline
\end{tabular}
\end{center}}
\end{table}

\begin{figure}[tbh]
\begin{center}
\hspace{0.5cm}
\psfig{file=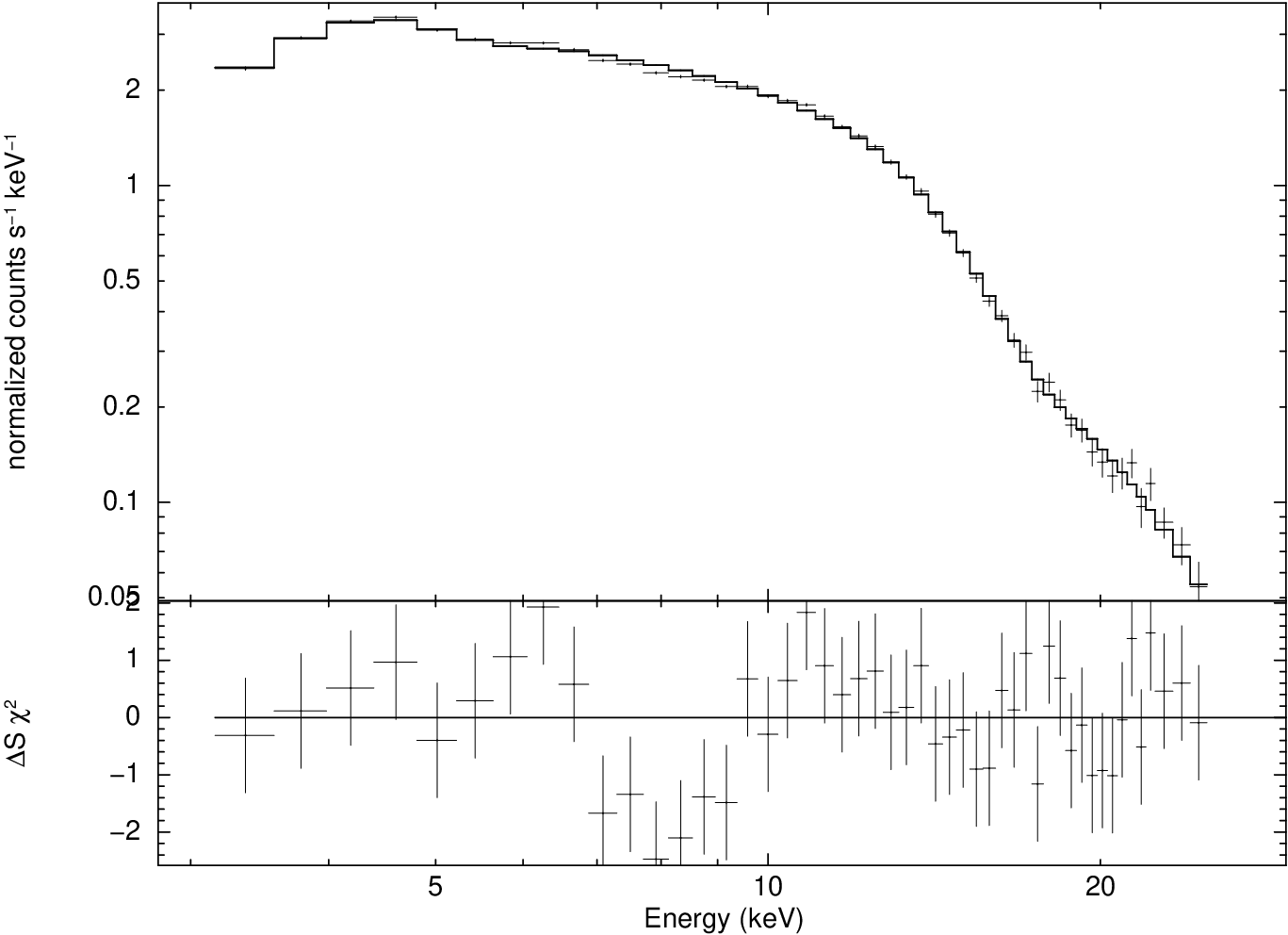,height=7cm,width=10cm,angle=-90}
\end{center}
\begin{center}
\small{Fig. 4 -- Average spectrum of all non-dip observations and the folded model is plotted. The bottom panel shows the residuals of the fit.}
\end{center}
\end{figure}

\section{Conclusion}

After more than 15 years long spin-down episode with an almost constant spin-down rate of  $\dot \nu=-3.54(2)\times 10^{-14}$ Hz s$^{-1}$ (Baykal et al.2001), RXTE observations between March 2001 and March 2002 showed a $\sim 0.6$ factor of decrease in spin down rate compared to long-term spin down rate of the source. From the observations of INTEGRAL between March 2003 and September 2005, Fritz et al. (2006) found that source was spinning-up with a rate of $2.58\times 10^{-14}$ Hz s$^{-1}$. We found, in this work, that the source began to spin-down again with a spin-down rate of $-3.59(2) \times 10^{-14}$ Hz s$^{-1}$ which is pretty close to the previous long term spin-down rate of the source.    

According to Ghosh\& Lamb (1979) model, net torque exerted on the neutron star by the accreting matter from a prograde accretion disk can be expressed as

\begin{equation}
N=I\dot{\Omega}=n(\omega_s)\dot{M}l,
\end{equation}  

where I is the moment of inertia of the neutron star, $\dot{\Omega}$ is the spin frequency derivative, $n(\omega_s)$ is the dimensionless torque, $\dot{M}$ is mass accretion rate and $l$ is specific angular momentum of the accreting matter. Here, $\omega_s$ is the fastness parameter defined as the ratio of neutron star's spin frequency to the Keplerian frequency at the magnetospheric radius, which is almost equal to the accretion disk's inner radius.

Dimensionless torque, $n(\omega_s)$ in Equation 2, is the ratio of the total (magnetic and material torque) to the material torque. It is, for a slowly rotating neutron star ($\omega_s<0.35-0.95$), expected to be positive and of the order of unity leading to spin up of the star, whereas it is expected to be negative for a fast rotating neutron star ($\omega_s>>0.35-0.95$) leading to spin down of the star (Ghosh\&Lamb 1979; Wang 1995; Li\&Wang 1996, 1999) . Since the dimensionless torque depends on the fastness parameter (and thus the magnetospheric radius), it decreases and may even be negative with an increase in magnetic field or decrease in mass accretion rate. Decrease in mass accretion rate leads to a decrease in X-ray flux and due to the accretion geometry changes, X-ray spectral variations may also be evident.

For 4U 1907+09, observed torque reversal cannot simply be explained by Ghosh\&Lamb (1979) model since spin-up episode observed by INTEGRAL and spin-down episodes with different spin-down rates have not shown significant X-ray flux and X-ray spectral variations (Baykal et al. 2001; Fritz et al. 2006; Baykal et al. 2006). Using this model, it is also difficult to explain similar spin rate magnitudes observed in spin-down and spin-up episodes of 4U 1907+09. Absolute value of the spin rate has always been found to be  $\sim 2-3\times 10^{-14}$ Hz s$^{-1}$. 

In case of accretion via stellar wind, simulations showed that transitions from spin-down to spin-up is possible even when there is not significant mass accretion rate change (Anzer et al. 1987;Taam\&Fryxell 1988a, 1988b, 1989; Blondin et al. 1990, Murray et al. 1998). However, torque reversal timescale of 4U 1907+09 is of the order of years which is much greater than the typical timescale of torque reversals in the simulations, found to be only of the order of hours. Moreover, 4U 1907+09 has shown transient QPOs (quasi-periodic oscillations) which is a sign of the presence of an accretion disk (in't Zand et al. 1998a; Mukerjee et al. 2001; Baykal et al. 2006).

It may also be possible that negative torques (corresponding to spin-down episodes) may
come from a retrograde Keplerian accretion disk (Nelson et al.
1997). Spin-down torques may also be the result of
an advection dominated sub-Keplerian disk for which the fastness
parameter should be higher than that of the rest of the
Keplerian disk causing a net spin-down (Yi et al. 1997) or the
warping of the disk so that the inner disk is tilted by more than
90 degrees (van Kerkwijk et al. 1998). Torque and X-ray luminosity
correlation is expected from these models which is not
found for 4U 1907+09.

A recent model of Perna et al. (2006) explains
torque changes in accreting neutron stars without a need for retrograde disks. In this model, the scenario in which the neutron star's magnetic field is tilted with respect to the axis of rotation of the neutron star is considered. For an accretion disk located in the neutron star's equatorial plane, the magnetic field strength depends
on the azimuthal angle and thus the location of magnetospheric radius
is variable. This can lead to regions in the disk where the propeller
effect is locally at work, while accretion from other regions is possible. Perna et al. (2006) shows that this accretion geometry may cause cyclic torque reversal episodes without a need for significant mass accretion rate variations.

4U 1907+09 has an eccentric orbit and shows occasional X-ray flux dips (in't Zand et al. 1997). These are indications of the transient accretion disk formation around the neutron star. On the other hand, model of Perna et al. (2006) is proposed for persistent prograde accretion disks like Ghosh\&Lamb (1979) model. To understand torque reversal of 4U 1907+09, transient nature of the accretion should also be taken into account.

From Figure 4, orbital dependence of $n_H$ is evident, on the other hand there is no significant variation of power law index. $n_H$ increases just after the periastron due to the accretion flow at the periastron passage and it remains at high values until the apastron. Roberts et al. (2001) also found similar $n_H$ dependence on orbital phase and interpreted this $n_H$ excess as the neutron star passing through an equatorially enhanced matter envelope in an inclined orbit. 

{\bf{Acknowledgments}}

S.\c{C}. \.{I}nam and A. Baykal acknowledge support from T\"{U}B\.{I}TAK, the Scientific and Technological Research Council of Turkey through project 106T040 and EU FP6 Transfer of Knowledge Project "Astrophysics of Neutron Stars"
(MTKD-CT-2006-042722). We thank Elif Beklen for useful discussions.

\noindent{{\bf{References}}}

Anzer U., B\"{o}rner G., Monaghan J., 1987, A\&A 176, 235

Baykal A., \.{I}nam S. \c{C}., Alpar M. A., et al. 2001, MNRAS, 327, 1269

Baykal A., \.{I}nam S. \c{C}., Beklen E. 2006, MNRAS, 369, 1760

Blondin J.M., Kalmann T.R., Fryxell B.A., et al. 1990, ApJ 356, 591

Chitnis V. R., Rao A. R., Agrawal P. C., Manchanda R. K. 1993, A\&A, 268, 609

Coburn W., Heindl W. A., Rothschild R. E. et al. 2002, ApJ, 580, 394

Cook M. C., Page C. G. 1987, MNRAS, 225, 381

Cox N. L. J., Kaper L.,  Mokiem M. R. 2005, A\&A, 436, 661

Cusumano G., di Salvo T., Burderi L., et al. 1998, A\&A, 338, L79

Deeter J.E., Boynton P.E. 1985, in Hayakawa S., Nagase F., eds, Proc. Inuyama
Workshop, Timing Studies of X-ray Sources, Nagoya Univ., Nagoya, p.29

Fritz S., Kreykenbohm I., Wilms J. et al. 2006, A\&A, 458, 885

Ghosh P., Lamb F.K. 1979, ApJ 234, 296

Giacconi R., Kellogg E., Gorenstein P., et al. 1971, ApJ, 165, L27

in't Zand J. J. M., Baykal A., Strohmayer T. E. 1998, ApJ, 496, 386

in't Zand J. J. M., Strohmayer T. E., Baykal A. 1997, ApJ, 479, L47

Iye M. 1986, PASJ, 38, 463

Jahoda K., Swank J., Giles A. B., et al. 1996, Proc. SPIE, 2808, 59

Leahy D.A., Darbro W., Elsner R.F., Weisskopf M.C., Sutherland P.G., Kahn S.,
Grindlay J.E. 1983, ApJ, 266, 160

Li X.-D., Wang Z.-R. 1996, A\&A 307, L5

Li, X.-D., Wang, Z.-R. 1999, ApJ, 513, L845

Makishima K., Kawai N., Koyama K. Shibazaki N. 1984, PASJ, 36, 679

Makishima K., Mihara T., Nagase F., Tanaka Y. 1999, ApJ, 525, 978

Makishima K., Ohashi T., Kawai N. et al., 1990, PASJ, 42, 295

Marshall N., Ricketts, M. J. 1980, MNRAS, 193, 7P

Miceli M., Decourchelle A., Ballet J. et al. 2006, A\&A, 453, 567

Mihara T. 1995, PhD thesis, RIKEN, Tokyo

Mukerjee K., Agrawal P. C., Paul B., et al. 2001, ApJ, 548, 368

Murray J.R., de Kool M., Li J. 1999, ApJ, 515, 738

Perna, R., Bozzo, E., Stella, L. 2006, ApJ, 639, 363

Roberts M. S. E., Michelson P. F., Leahy D. A., et al. 2001, ApJ, 555, 967

Schwartz D. A., Griffiths R. E., Thorstensen J. R., et al.1980, AJ, 85, 549

Taam R.E., Fryxell B.A. 1988a, ApJ 327, L73

Taam R.E., Fryxell B.A. 1988b, ApJ 335, 862

Taam R.E., Fryxell B.A. 1989, ApJ 339, 297

Valinia A., Marshall F. E., 1998, ApJ, 505, 134

van Kerkwijk M. H., van Oijen J. G. J., van den Heuvel E. P. J. 1989, A\&A, 209, 173

van Kerkwijk M. H., Chakrabarty D., Pringle J. E., Wijers R. A. M. J. 1998, ApJ, 499, L27

Yi I., Wheeler J.C., Vishniac E.T. 1997, ApJ 481, L51

Wang Y.-M. 1995, ApJ 449, L153

\end{document}